\documentclass[preprint,12pt]{elsarticle}

\usepackage{amssymb}

\journal{Physics Letters A}

\def\p{\partial}
\def\a{\alpha}

\def\lb{\lambda}

\def\vfi{\varphi}

\def\dg{\dagger}

\def\cD{{\cal D}}

\newcommand{\Sc}{Schr\"odinger }
\newcommand{\eqref}[1]{(\ref{#1})}

\newcommand{\be}[1]{\begin{equation}\label{#1}}
\newcommand{\ee}{\end{equation}}
\newcommand{\ba}[1]{\begin{eqnarray}\label{#1}}
\newcommand{\ea}{\end{eqnarray}}
\newcommand{\rf}[1]{(\ref{#1})}

\newcommand{\ra}{\rangle}
\newcommand{\la}{\langle}

\begin{document}

\begin{frontmatter}

\title{Equivalent Hermitian operator from supersymmetric quantum
mechanics}

\author[rvt]{Boris F. Samsonov}
\ead{samsonov@phys.tsu.ru}
\author[rvt,focal]{V. V. Shamshutdinova\corref{cor2}}
\ead{shvv@phys.tsu.ru}
\author[rvt]{A. V. Osipov}
\ead{avopiso@phys.tsu.ru}  \cortext[cor2]{Corresponding author}

\address[rvt]{Physics Department, Tomsk State University,
36 Lenin Avenue, 634050 Tomsk, Russian Federation}
\address[focal]{Natural Science and
Mathematics Department, Tomsk Polytechnic University, 30 Lenin
Avenue, 634050 Tomsk, Russian Federation}

\begin{abstract}
Diagonalizable pseudo-Hermitian Hamiltonians with real and
discrete spectra, which are superpartners of Hermitian
Hamiltonians, must be $\eta$-pseudo-Hermitian with Hermitian and
invertible $\eta$ operators. We show that despite the fact that
an $\eta$ operator produced by a supersymmetric transformation,
corresponding to the exact supersymmetry,
is not invertible in the whole Hilbert space, it can be
used to find the eigenfunctions of a Hermitian operator equivalent
to the given pseudo-Hermitian Hamiltonian. Once the eigenfunctions
of the Hermitian operator are found the operator may be
reconstructed with the help of the spectral decomposition.
\end{abstract}


\end{frontmatter}

After works by Andrianov and Cannata with coworkers
\cite{Andr-Ioffe} the supersymmetric quantum mechanics (SUSY QM)
became an effective tool in studying different properties of the
\Sc equation with a complex-valued potential. Thus in
\cite{S-irred} the author shows that new exactly solvable
complex-valued potentials may be obtained with the help of second
order irreducible SUSY transformations. In \cite{S-d-n} a
possibility is discussed to transform non-diagonalizable
Hamiltonians into diagonalizable ones and vice versa and in
\cite{S-ss} the author stresses that SUSY transformations may
create Hamiltonians with spectral singularities inside the
continuous spectrum. Green functions for some complex SUSY
partners of real potentials are calculated in \cite{S-P-PLA}. In
\cite{GSF} SUSY QM is applied to study a dynamo effect and in
\cite{DDT} spontaneous breakdown of {\cal PT} symmetry is related
with the presence of the supersymmetry in the system.

One of the main features of supersymmetric transformations
corresponding to the exact supersymmetry is that the
transformation operator has a nontrivial kernel. This
results in the property that a non-Hermitian diagonalizable
Hamiltonian with a real and discrete spectrum, which is a
super-partner of an initial Hermitian Hamiltonian, becomes
(weak) $\eta$-pseudo-Hermitian where $\eta=\eta^\dg$ is not
invertible in the whole Hilbert space. The possibility for
$\eta$ to be non-invertible was first considered by Fitio
\cite{Fityo} and latter explored in \cite{MM} in the
context of generating non-Hermitian Hamiltonians with real
spectra. Nevertheless, these authors completely skipped the
question whether or not a new Hilbert space, where the
non-Hermitian operator becomes Hermitian, may be
constructed with the help of such an $\eta$ operator.
Indeed, if $\eta=o^\dg o$ has a nontrivial kernel,
$\mbox{ker}\,o\ne\O$,
a new inner product defined as (see e.g. \cite{Mostafaz1})
\be{1}
\langle\langle\psi,\phi\rangle\rangle:=
\langle\psi|\eta|\phi\rangle\,,\quad\eta=\eta^\dg
\ee
becomes positive-semidefinite even if the Hamiltonian is
diagonalizable. In view of a theorem proved in
\cite{Mostafaz2} this means that such an $\eta$ operator is
not suitable for constructing the new inner product and
there should exist another $\eta$ operator which is
invertible in the whole Hilbert space.

The main aim of the current Letter is to show that such an $\eta$
operator may be used for constructing a Hermitian operator
equivalent to the given non-Hermitian one acting in the same
Hilbert space. Thus, the approach based on equivalence classes
(see \cite{GS}) proves to be more advantageous in this case than
the usual scheme  \cite{Mostafaz1,Mostafaz2} based on Eq.
\eqref{1}.

For simplicity we will consider a regular Sturm-Liouville problem
defined by a differential expression $h=-\p_x^2+V_0(x)$
($\p_x\equiv\frac{d}{dx}$) with a real-valued potential function
$V_0(x)$, $x\in[0,d]$ and the Dirichlet boundary conditions at
$x=0,d$. Under rather mild conditions on the function $V_0(x)$
($V_0(x)$ is real and has a bounded derivative, see e.g.
\cite{KS}) this differential equation together with the Dirichlet
boundary conditions defines a selfadjoint operator in the Hilbert
space $L^2(0,d)$ of square integrable functions over the interval
$[0,d]$. To simplify notations we will denote this operator by the
same symbol $h=h^\dg$. Then the corresponding boundary value
problem
 \be{bvph}
h\psi=E\psi\,,\quad \psi(0)=\psi(d)=0
 \ee
(and, hence, operator $h$) has an unbounded real and discrete
spectrum \cite{KS} $E=E_n=E_n^*$ with eigenfunctions
$\psi=\psi_n=\psi_n^*$, $n=1,2,\ldots$

Let $D_h$ be the domain of definition of $h$. It consists of the
twice continuously differentiable functions defined $\forall
x\in(0,d)$ satisfying the Dirichlet boundary conditions. Let $L$
be a first order differential intertwining operator between $h$
and a non-selfadjoint operator $H=-\p_x^2+V(x)$
 \be{LhH}
Lh=HL
 \ee
where in general $V(x)$ is a complex-valued function which we will
describe below. Actually, for the moment we consider $H$ only as a
differential expression but below we will supply it with proper
boundary conditions which will transform $H$ into a
non-selfadjoint operator acting in the same Hilbert space
$L^2(0,d)$. Operator $L$ is defined in terms of a complex-valued
superpotential $W=W(x)$ as follows
 $L=-\p_x+W$ where $W=u_x/u$
($u_x:=du(x)/dx$) with $u(x)$ being an (essentially) complex (i.e.
$W$ must be complex) and nodeless solution to the differential
equation $hu=\alpha u$, $\a\in\Bbb R$ and $Lu=0$ (see e.g.
\cite{Andr-Ioffe}). Note that if $\a$ coincides with one of the
levels $E_{n}$, $n=1,2,\ldots$, i.e. $\a=E_{n_0}$, the Hamiltonian
$H$ becomes non-diagonalizable (see e.g. \cite{S-d-n}), a
possibility we would like to avoid. Therefore we will assume
$\a\ne E_n$, $n=1,2,\ldots$. If now $\psi=\psi(x)$ is a solution
to the differential equation $(h-E)\psi=0$ then
$\vfi=\vfi(x)=L\psi(x)$ satisfies differential equation
$(H-E)\vfi=0$, $H=-\p_x^2+V(x)$, $V(x)=V_0(x)-2W_x$.
By this reason we
have
 \be{vfi}
 \vfi_x+W\vfi=(E-V_0+W_x+W^2)\psi\,.
 \ee
From \eqref{vfi} it follows that if $\psi$ satisfies the Dirichlet
boundary conditions (see \rf{bvph}) then
\be{Hbc}
(\vfi_x+W\vfi)(0)=(\vfi_x+W\vfi)(d)=0\,.
\ee
From here we conclude
that the functions $\vfi_n=L\psi_n$, $n=1,2,\ldots$ solve the
following boundary value problem
\be{bvpH}
H\vfi_n=E_n\vfi_n\,,\quad
(\vfi_{nx}+W\vfi_n)(0)=(\vfi_{nx}+W\vfi_n)(d)=0\,.
\ee
This means that in the Hilbert space $L^2(0,d)$ a non-selfadjoint operator
$H$ with the domain of definition $D_H$ is defined. Domain $D_H$
consists of the twice continuously differentiable functions
defined $\forall x\in(0,d)$ satisfying boundary conditions
\eqref{Hbc}. It is important to note that the function
$\vfi_0=1/u$ also solves the boundary value problem \eqref{bvpH}.
This means that the spectrum of $H$ consists of the levels $E_n$,
$n=1,2,\ldots$ plus $E=E_0=\a\ne E_n$. From the point of view of
SUSY QM this corresponds to the exact supersymmetry (see e.g.
\cite{CKS}).

In what follows we do not need to use the full domains of
definition of the operators $h$ and $H$. Instead we will use
domains $\cD_h$ and $\cD_H$ which are the sets of all finite
linear combinations of the eigenfunctions of $h$ and $H$
respectively. Operator $L$ is well defined on $\cD_h$ and maps
$\cD_h\to\cD_H$.

Operator $H^\dg$ Hermitian adjoint to $H$ is defined by the
adjoint boundary value problem
\be{bvpHd}
H^\dg\xi=E\xi\,,\quad
(\xi_{x}+W^*\xi)(0)=(\xi_{x}+W^*\xi)(d)=0
 \ee
with $H^\dg=-\p_x^2+V^*(x)$. Its domain of definition $D_{H^\dg}$
consists of the  twice continuously differentiable functions
defined $\forall x\in(0,d)$ with the boundary conditions as given
in \eqref{bvpHd}. The spectrum of $H^\dg$ coincides with the
spectrum of $H$, $E=E_n$, $n=0,1\ldots$. Its eigenfunctions
$\xi=\xi_n=\vfi_n^*$, $n=0,1,\ldots$ form a basis in $L^2(0,d)$
biorthogonal to $\{\vfi_n\}$
\be{biort}
\la\xi_n|\vfi_m\ra=0\,,\quad n\ne m\,.
\ee
In what follows we will use $\cD_{H^\dg}$ which is the set of all finite linear
combinations of ${H^\dg}$ eigenfunctions $\xi_n$.

From \eqref{LhH} it follows that
\be{hLd}
hL^\dg=L^\dg H^\dg
\ee
where $L^\dg=\p_x+W^*$. This relation means that $L^\dg$
transforms eigenfunctions of $H^\dg$ to the eigenfunctions of $h$
except for $\xi_0=\vfi^*_0=1/u^*$ which enters the kernel of
$L^\dg$, $L^\dg\xi_0=0$. Operator $L^\dg$ is well defined on
$\cD_{H^\dg}$. It realizes the mapping $\cD_{H^\dg}\to\cD_h$.

From \eqref{LhH} and \eqref{hLd} one deduces that
\be{HLLd}
HLL^\dg=LL^\dg H^\dg\,.
\ee
Since the spectrum of $H$ is real, it
should be $\eta$-pseudo-Hermitian \cite{Mostafaz2}, i.e. there
should exist an invertible Hermitian operator $\eta$ such that
$H^\dg=\eta H\eta^{-1}$ or equivalently
\be{etaH}
\eta H=H^\dg\eta\,.
\ee
If $LL^\dg$ were invertible, comparing
\rf{HLLd} and \rf{etaH} we could conclude that
$\eta=(LL^\dg)^{-1}$. Unfortunately this is not so. Nevertheless,
as we show below, the operator $LL^\dg$ can be used to calculate
eigenfunctions of a Hermitian operator $h_0=h_0^\dg$ equivalent to
$H^\dg$ (and, hence, to $H$). Then $h_0$ may be reconstructed with
the help of the spectral decomposition. Note that composition
$LL^\dg$ is well defined on $\cD_{H^\dg}$ and maps
$\cD_{H^\dg}\to\cD_{H}$ while its domain of definition is
$D_{H^\dg}$.

Operator $LL^\dg$ is a second order differential operator,
\be{LLd}
LL^\dg=-\p_x^2+(W-W^*)\p_x+WW^*-W^*_x
\ee
and in the space $\cD_{H^\dg}$ it has a one-dimensional kernel with the basis
function $\vfi_0^*=1/u^*$. From \rf{HLLd} it follows that it
transforms eigenfunctions of $H^\dg$ to eigenfunctions of $H$
except for $E=\a$. Therefore using biorthogonality relation
\rf{biort} one gets
\be{spLLd}
\la\xi_n|LL^\dg\xi_m\ra=0\,,\quad
n\ne m\,.
\ee
It is not difficult to see that being defined on
$D_{H^\dg}$ operator $LL^\dg$ is selfadjoint and positive
semidefinite. Therefore the boundary value problem
\be{bvpLLd}
LL^\dg\Xi=\lb^2\Xi\,,\quad
(\Xi_{x}+W^*\Xi)(0)=(\Xi_{x}+W^*\Xi)(d)=0
 \ee
has a real non-negative spectrum $\lb^2=\lb_n^2$, $n=0,1,\ldots$,
$\lb_0=0$ with eigenfunctions $\Xi=\Xi_n$, $n=0,1,\ldots$,
$\Xi_0=\vfi^*_0$.
Its Hermitian square root is well defined in
$L^2(0,d)$ and may be written using its spectral decomposition
\be{LLdsr}
(LL^\dg)^{1/2}=\sum_{k=0}^\infty\lb_k|\Xi_k\ra\la\Xi_k|\,.
\ee

Let us consider the functions
\be{LLdxi}
\Phi_n=(LL^\dg)^{1/2}\xi_n\,,\quad n=1,2,\ldots, \quad
\Phi_0=\Xi_0\,.
\ee
From \rf{spLLd} it follows that
$\la\Phi_n|\Phi_m\ra=0$, $n\ne m$, $n,m=1,2\ldots$\,.
 Moreover,
$\la\Phi_n|\Phi_0\ra=0$, $n=1,2\ldots$ since $LL^\dg\Xi_0=0$.
Thus the functions $\Phi_n$ \rf{LLdxi} form an orthogonal set.
Furthermore, since $\xi_n$, $n=0,1,\ldots$ form a basis in
$L^2(0,d)$ the functions \rf{LLdxi} form an orthogonal basis in
$L^2(0,d)$.

Denote $\cD_{h_{01}}$ the space of all finite linear
combinations of the functions $\Phi_n$, $n=1,2,\ldots$.
Being
restricted to this space, $(LL^\dg)^{1/2}$
becomes invertible.
Therefore operator
\be{h01}
h_{01}=(LL^\dg)^{1/2}H^\dg(LL^\dg)^{-1/2}
\ee
is well defined on
$\cD_{h_{01}}$.
Moreover, from intertwining relation \rf{HLLd} it
follows that it is Hermitian on $\cD_{h_{01}}$,
$h_{01}=h_{01}^\dg$ and $h_{01}\Phi_n=E_n\Phi_n$, $n=1,2,\ldots$.
Therefore Eq. \rf{h01} may be rewritten as a couple of
intertwining relations
\be{h01i}
(LL^\dg)^{1/2}h_{01}=H(LL^\dg)^{1/2}\,,\quad
(LL^\dg)^{1/2}H^\dg=h_{01}(LL^\dg)^{1/2}\,.
\ee
Eqs. \rf{h01i} mean that the operator $(LL^\dg)^{1/2}$, being
restricted to $\cD_{h_{01}}$, realizes an equivalence
transformation between a restriction of $H$ and
$h_{01}$.
More precisely it transforms eigenfunctions $\Phi_n$
of $h_{01}$ to eigenfunctions $\vfi_n$ of $H$,
$(LL^\dg)^{1/2}\Phi_n=\vfi_n$, $n=1,2,\ldots$.
Similarly, since $(LL^\dg)^{1/2}$ transforms eigenfunctions
$\xi_n$ of $H^\dg$ to eigenfunctions $\Phi_n$ of $h_{01}$,
$(LL^\dg)^{1/2}\xi_n=\Phi_n$, $n=1,2,\ldots$,
it realizes an equivalence transformation between $h_{01}$
and a restriction of $H^\dg$.

 Now we will extend operator $h_{01}$ from the
space $\cD_{h_{01}}$ to the space $\cD_{h_0}$ of all finite linear
combinations of the functions $\Phi_n$, $n=0,1,\ldots$. By
construction one has $\cD_{h_0}=\cD_{h_{01}}\oplus\cD_{h_{00}}$
where $\cD_{h_{00}}$ is a one-dimensional space with the basis
function $\Phi_0$ which is orthogonal to $\cD_{h_{01}}$.
By this
reason if we fix the additional eigenvalue of $h_0$ to be equal to
$E_0=\a$, this continuation,
if restricted to be selfadjoint,
 becomes unique
\be{h0}
h_0=\sum_{n=0}^\infty |N_n|^2E_n|\Phi_n\ra\la\Phi_n|\,.
\ee
Here
$N_n$ is a normalization coefficient of the function $\Phi_n$.
Although this approach permitted us to construct a unique operator
$h_0$ equivalent to $H^\dg$, an invertible $\eta$ operator may
lead to another $h_0$ which is unitary equivalent to \rf{h0}.

Let us consider the simplest case when the boundary value problem
\rf{bvpLLd} may be solved exactly. It corresponds to
transformation function $u=\exp{(iax)}$ with $W=ia$, $\a=a^2$. The
initial boundary value problem \rf{bvph} in this case has zero
potential. For the transformation operator one gets $L=-\p_x+ia$
and $LL^\dg=-\p_x^2+2ia\p_x+a^2$. The spectrum and eigenfunctions of $LL^\dg$ are
\be{lbn2} \lb_n^2=k_n^2\,,\quad k_n=\pi n/d\,,\quad n=0,1,2,\ldots
\ee
\[
\Xi_n=\sqrt{\frac2d}\,e^{iax}\cos(k_nx)\,,\quad n=1,2,\ldots,\quad
\Xi_0=\sqrt{\frac1d}\,e^{iax}\,.
\]

Since $W$ does not depend on $x$ the operator $H$ has still zero
potential but it is non-selfadjoint because of boundary conditions
\rf{Hbc} which in this particular case read
\be{RBC}
(\vfi_x+ia\vfi)(0)=(\vfi_x+ia\vfi)(d)=0\,.
\ee
We recognize here the boundary value problem generated by the Robin boundary
conditions previously studied in \cite{Czhech1,Czhech2}. The
spectrum of $H$ contains all energy levels corresponding to the
boundary value problem \rf{bvph} with zero potential $E_n=k_n^2$,
$n=1,2,\ldots$ plus the level $E_0=\a=a^2$ and the supersymmetry
is exact. The corresponding eigenfunctions (not normalized to
unity) read
\[
\vfi_n=\cos(k_nx)-\frac{ia}{k_n}\sin(k_nx)\,,\quad n=1,2,\ldots
,\quad \vfi_0=e^{-ia x}\,.
\]
Eigenfunctions of $H^\dg$ are simply complex conjugated
eigenfunctions of $H$, $\xi_n=\vfi_n^*$. As it was mentioned above
if $\a=E_n$, $n=1,2,\ldots$ both Hamiltonian $H$ and $H^\dg$
become non-diagonalizable and therefore we have to assume $a\ne\pi
n/d$, $n=1,2,\ldots$.

To find eigenfunctions $\Phi_n$, $n=1,2,\ldots$ according to
\rf{LLdsr} and \rf{LLdxi} we have to evaluate integrals
$s_{mn}=\la\Xi_m|\xi_n\ra$, which are almost standard (see e.g.
\cite{Prudnikov}) \be{smn} s_{mn}=\frac{2ia d\sqrt{2d}e^{-ia
d}((-1)^{m+n}-e^{ia d})
(a^2d^2-m^2\pi^2)}{a^4d^4-2a^2d^2(m^2+n^2)\pi^2+(m^2-n^2)^2\pi^4}
\ee and then to sum up the series \rf{LLdsr} with $(LL^\dg)^{1/2}$
applied to $\xi_n$. Using Eqs. \rf{lbn2} and \rf{smn} one can
reduce this series to a combination of two known series (see
\cite{Prudnikov})
\begin{equation}\label{}
\sum_{n=0}^\infty\frac{\cos
nx}{n+\rho}=\beta(\rho)\cos[(\pi-x)\rho]+
\frac{1}{2}\int_x^\pi\cos[(\rho-1/2)t-x\rho]\csc\frac{t}{2}dt,
\end{equation}
\begin{equation}\label{}
\sum_{n=0}^\infty\frac{(-1)^n\cos
nx}{n+\rho}=\beta(\rho)\cos[x\rho]-
\frac{1}{2}\int_0^x\sin[(\rho-1/2)t-x\rho]\sec\frac{t}{2}dt\,.
\end{equation}
Finally for the eigenfunctions of $h_0$ we get
\begin{eqnarray*}
\Phi_0&=&e^{iax},\\
\Phi_n&=&\frac{e^{i(x-d)a}\left(a^2d^2-n^2\pi^2\right)}{2nd\pi}\\
&\times&\left\{\frac{e^{iad}}{\pi}\left(\gamma(\zeta)\cos\left[\pi\zeta-\frac{x\pi\zeta}{d}\right]-
\gamma(\delta)\cos\left[\pi\delta-\frac{x\pi\delta}{d}\right]\right)\right.\\
&+&\left.\frac{e^{iad}}{\pi}\int_\frac{\pi
x}{d}^\pi\left(\cos\left[\frac{\pi
x\zeta}{d}-t\zeta\right]-\cos\left[\frac{\pi
x\delta}{d}-t\delta\right]\right)\cot\frac{t}{2}dt\right.\\
&+&\left.\frac{(-1)^{n}}{\pi}\left(\gamma(\delta)\cos \frac{\pi
x\delta}{d}-\gamma(\zeta)\cos
\frac{\pi x\zeta}{d}\right)\right.\\
&+&\left.\frac{(-1)^{n}}{\pi}\int_0^\frac{\pi
x}{d}\left(\cos\left[\frac{\pi
x\delta}{d}-t\delta\right]-\cos\left[\frac{\pi
x\zeta}{d}-t\zeta\right]\right)\tan\frac{t}{2}dt\right\}.
\end{eqnarray*}
Here $n=1,2\ldots$, $\delta\equiv \frac{ad}{\pi}-n$,
$\zeta\equiv\frac{ad}{\pi}+n$ and
$$\gamma(z)=\beta\left(z\right)+\beta\left(-z\right),\quad
\beta\left(z\right)=\frac{1}{2}\left(\frac{\Gamma'[\frac{z+1}{2}]}%
{\Gamma[\frac{z+1}{2}]}-\frac{\Gamma'[\frac{z}{2}]}{\Gamma[\frac{z}{2}]}\right)$$
with $\Gamma[z]$ being the Euler gamma function.

To summarize, we have shown that despite the fact that the operator $LL^\dag$
produced by SUSY QM with exact supersymmetry is not
invertible, it can be used to find eigenfunctions of Hermitian
operator equivalent to non-Hermitian operator generated by SUSY
QM. Once these eigenfunctions are found the Hermitian operator may
be reconstructed using its spectral decomposition.

Acknowledgment. BFS is grateful to A. Mostafazadeh for drawing his
attention to paper \cite{Czhech1}. The authors would like to thank
A.V. Sokolov for useful comments. The work is partially supported
by President of Russia under the grant SS-871.2008.2, Russian
Science and Innovations Federal Agency under contract No
02.740.11.0238 and Russian Federal Agency of Education under
contract No P1337 and No P2596.


\begin{thebibliography}{00}
\bibitem{Andr-Ioffe} A.A. Andrianov, F. Cannata, J.-P. Dedonder, M.V. Ioffe, Int. J. Mod. Phys. A 14 (1999)
2675;\\
F. Cannata, G. Junker, J. Trost,
Phys. Lett. A 246 (1998) 219;\\
 F. Cannata, M. Ioffe, R. Roychoudhury, P. Roy,
Phys. Lett. A 281 (2001) 305;\\
 A.A. Andrianov, F. Cannata, A.V. Sokolov,
Nucl. Phys. B 773 (2007) 107;\\
 A.V. Sokolov,
Nuclear Phys. B 773 (2007) 137.

\bibitem{S-irred} B.F. Samsonov,
Phys. Lett. A 358 (2006) 105;

\bibitem{S-d-n} B.F. Samsonov, J. Phys. A: Math. Gen. 38 (2005) L397;\\
B.F. Samsonov, P. Roy, J. Phys. A: Math. Gen. 38 (2005) L249.

\bibitem{S-ss}
B.F. Samsonov, J. Phys. A: Math. Gen. 38 (2005) L571.

\bibitem{S-P-PLA} B.F. Samsonov, A.M. Pupasov, Phys. Lett. A 356 (2006) 210.

\bibitem{GSF} U. Guenther, B.F. Samsonov, F. Stefani,
   J. Phys. A: Math. Theor. 40 (2007) F169.

\bibitem{DDT} P. Dorey, C. Dunning, R. Tateo, J. Phys. A: Math. Gen. 34
(2001) L391.

\bibitem{Fityo}T.V. Fityo, J. Phys. A: Math. Gen. 35 (2002) 5893.


\bibitem{MM}O. Mustafa, S.H. Mazharimousavi, Phys. Lett. A 357 (2006) 295;\\
O. Mustafa, S.H. Mazharimousavi, J. Phys. A: Math. Theor. 41 (2008) 244020;\\
O. Mustafa, S.H. Mazharimousavi, Int. J. Theor. Phys. 47 (2008)
2029.


\bibitem{Mostafaz1}
A. Mostafazadeh, J. Math. Phys. 43 (2002) 6343.

\bibitem{Mostafaz2}
A. Mostafazadeh, J. Math. Phys. 43 (2002) 2814.

\bibitem{GS}
U. Gunther, B.F. Samsonov, Phys. Rev. A 78 (2008) 042115.

\bibitem{KS}
B.M. Levitan, I.S. Sargsyan, Introduction to spectral theory:
Selfadjoint ordinary differential operators. Translations of
mathematical monographs, 39, AMS, Providence, Rhode Island, 1975.

\bibitem{CKS}
 F. Cooper, A.A. Khare, U. Sukhatme, Phys. Rep. 251 (1995) 267.

\bibitem{Czhech1}D. Krej\v{c}i\v{r}\'{i}k, H.
B\'{i}la, M. Znojil, J. Phys. A: Math. Gen. 39 (2006) 10143.

\bibitem{Czhech2}D. Krej\v{c}i\v{r}\'{i}k, J. Phys. A: Math. Theor. 41 (2008)
244012.

\bibitem{Prudnikov} A.P. Prudnikov, U.A. Brychkov, O.I. Marichev,
Integrals and series, Nauka, Moscow, 1981, (in Russian).


\end{thebibliography}
\end{document}